\documentclass[copyright,creativecommons]{eptcs}
\providecommand{\event}{EXPRESS/SOS 2015}

\usepackage[ngerman,english]{babel}
\usepackage[utf8x]{inputenc}
\usepackage{amsmath,amsfonts,amssymb,amsthm}
\usepackage{stmaryrd}
\usepackage{extarrows}
\usepackage{multirow}
\usepackage{xspace}
\usepackage{paralist}
\usepackage{graphicx}
\usepackage{color}
\usepackage{tikz}
\usetikzlibrary{shapes.geometric}
\usetikzlibrary{calc,positioning}

\usepackage{count1to}
\usepackage{semantic}
\usepackage{url}
\usepackage[colorinlistoftodos]{todonotes}
\usepackage{cite}
\usepackage{bigdelim}
\usepackage{etoolbox}
\usepackage{hyperref}
\usepackage{cleveref}

\usepackage{macros}

\title{Encoding CSP into CCS
	\thanks{Supported by the DFG via project ``Synchronous and Asynchronous Interaction in Distributed Systems".}}

\author{Meike Hatzel 
	\institute{TU Berlin}
	\and Christoph Wagner
	\institute{TU Berlin}
	\and Kirstin Peters\thanks{Supported by the German Federal and State Governments via the Excellence Initiative (Institutional Strategy).}
	\institute{TU Dresden}
	\and Uwe Nestmann
	\institute{TU Berlin}
}
\def\titlerunning{Encoding CSP into CCS}
\def\authorrunning{M.\ Hatzel, C.\ Wagner, K.\ Peters, U.\ Nestmann}
\def\copyrightholders{Hatzel, Wagner, Peters \& Nestmann}

\begin{document}

\maketitle

\begin{abstract}
	We study encodings from CSP into asynchronous CCS with name passing and matching, so in fact, the asynchronous $\pi$-calculus. By doing so, we discuss two different ways to map the multi-way synchronisation mechanism of CSP into the two-way synchronisation mechanism of CCS. Both encodings satisfy the criteria of Gorla except for compositionality, as both use an additional top-level context. Following the work of Parrow and Sjödin, the first encoding uses a centralised coordinator and establishes a variant of weak bisimilarity between source terms and their translations. The second encoding is decentralised, and thus more efficient, but ensures only a form of coupled similarity between source terms and their translations.
\end{abstract}

\section{Introduction}

In the context of a scientific meeting on Expressiveness in Concurrency and Structural Operational Semantics (SOS), likely very little needs to be said about the process algebras (or process calculi) CSP and CCS. Too many papers have been written since their advent in the 70's to be mentioned in our own paper; it is instructive, though, and recommended to appreciate Jos Baeten's historical overview \cite{Baeten:2005:BHP:1085667.1085669}, which also places CSP and CCS in the context of other process algebras like ACP and the many extensions by probabilities, time, mobility, etc. Here, we just select references that help to understand our motivation.

\vspace{0.3em}
\noindent
\textbf{Differences.}\;
From the beginning, although CSP \cite{hoare:78csp} and CCS \cite{CCS} were intended to capture, describe and analyse reactive and interactive concurrent systems, they were designed following rather different philosophies. Tony Hoare described this nicely in his position paper \cite{Hoare2006209} as follows: ``A primary goal in the original design of CCS was to discover and codify a minimal set of basic primitive agents and operators \dots and a wide range of useful operators which have been studied subsequently are all definable in terms of CCS primitives." and ``CSP was more interested in this broader range of useful operators, independent of which of them might be selected as primitive." So, at their heart, the two calculi use two different synchronisation mechanisms, one (CCS) using binary, \ie two-way, handshake via matching actions and co-actions, the other (CSP) using multiway synchronisation governed by explicit synchronisation sets that are typically attached to parallel composition. Another difference is the focus on Structural Operational Semantics in CCS, and the definition of behavioural equivalences on top of this, while CSP emphasised a trace-based denotational model, enhanced with failures, and the question on how to design models such that they satisfy a given set of laws of equivalence.

\vspace{0.3em}
\noindent
\textbf{Comparisons.}\;
From the early days, researchers were interested in more or less formal comparisons between CSP and CCS. This was carried out by both Hoare \cite{Hoare2006209} and Milner \cite{DBLP:conf/ifip/Milner86} themselves, where they concentrate on the differences in the underlying design principles. But also other researchers joined the game, but with different analysis tools and comparison criteria. 

For example, Brookes \cite{DBLP:conf/icalp/Brookes83} contributed a deep study on the relation between the underlying abstract models, synchronisation trees for CCS and the failures model of CSP. Quite differently, Lanese and Montanari \cite{Lanese200655} used the power to transform graphs as a measure for the expressiveness of the two calculi. 

Yet completely differently, Parrow and Sj{\"o}din \cite{sjodin:phd,parrowCoupled92} tried to find an algorithm to implement|best in a fully distributed fashion|the multiway synchronisation operator of CSP (and its variant LOTOS \cite{DBLP:conf/pstv/Brinksma85}) using the supposedly simpler two-way synchronisation of CCS. They came up with two candidates|a reasonably simple centralised synchroniser, and a considerably less simple distributed synchroniser\footnote{Recently \cite{7092761}, a slight variant of the protocol behind this algorithm was used to implement the distributed compiler DLC for a substantial subset of LNT (successor of LOTOS New Technology) that yields reasonably efficient C code.}|and proved that the two are not weakly bisimilar, but rather coupled similar, which is only slightly weaker. Coupled simulation is a notion that Parrow and Sj\"odin invented for just this purpose, but it has proved afterwards to be often just the right tool when analysing the correctness of distribution- and divergence-sensitive encodings that involve partial commitments (whose only effect is to gradually perform internal choices) \cite{nestmannPierce00}.

The probably most recent comparison between CSP and CCS was provided by van Glabbeek~\cite{DBLP:journals/corr/abs-1208-2750}. As an example for his general framework to analyse the relative expressive power of calculi, he studied the existence of syntactical translations from CSP into CCS, for which a common semantical domain is provided via labelled transition systems (LTS) derived from respective sets of SOS rules. The comparison is here carried out by checking whether a CSP term and its translation into CCS are distinguishable with respect to a number of equivalences defined on top of the LTS. The concrete results are: (1)~there is a translation that is correct up to trace equivalence (and contains deadlocks), and (2)~there is no translation that is correct up to weak bisimilarity equivalence that also takes divergence into account.

\vspace{0.3em}
\noindent
\textbf{Contribution.}\;
Given van Glabbeek's negative result, and given Parrow and Sj\"odin's algorithm, we set out to check whether we can define a syntactical encoding from CSP into CCS|using Parrow and Sj\"odin's ideas|that is correct up to coupled similarity.\footnote{The idea and a first draft of the encoding were developed by Nestmann and van Glabbeek during a stay at NICTA, Sydney.} We almost managed. In this paper, we report on our current results along these lines: 
(1)~Our encoding target is an asynchronous variant of CCS, but enhanced with name-passing and matching, so it is in fact an asynchronous $\pi$-calculus; we kept mentioning CCS in the title of this paper, as it clearly emphasises the origin and motivation of this work. But, we could \emph{not} do without name-passing.
(2)~We exhibit one encoding that is not distributability-preserving (so, it represents a centralised solution), but is correct up to weak bisimilarity and does not introduce divergence. This does not contradict van Glabbeek's results, but suggests that van Glabbeek's framework implies some form of distributability-preservation.
(3)~We exhibit another encoding that \emph{is} distributability-preserving and divergence-reflecting, but is only correct up to coupled similarity.

\vspace{0.3em}
\noindent
\textbf{Overview.}\;
We introduce the considered variants of CSP and CCS in \S~\ref{sec:techPrel}. There we also introduce the criteria---that are (variants of) the criteria in \cite{gorla10} and \cite{petersNestmannGoltz13}---modulo which we prove the quality of the considered encodings. In \S~\ref{sec:innerPart} we introduce the inner layer of our two encodings. It provides the main machinery to encode synchronisations of CSP. We complete this encoding with an outer layer that is either a centralised (\S~\ref{sec:central}) or a decentralised coordinator (\S~\ref{sec:decentral}). In \S~\ref{sec:conclusion} we discuss the two encodings. Missing proofs and some additional informations can be found in \cite{hatzelTechRep15}.

\section{Technical Preliminaries}
\label{sec:techPrel}

A process calculus $ \left( \proc, {\step} \right) $ consists of a set $ \proc $ of processes (syntax) and a reduction relation $ {\step} \subseteq \proc^2 $ (semantics).
Let $\Names$ be the countably-infinite set of names.
$\tau \not\in \Names$ denotes an internal unobservable action.
We use $ a, b, x, \ldots $ to range over names and $ P, Q, \ldots $ to range over processes.
We use $\alpha, \beta \ldots$ to range over $\Names \cup \{\tau\}$.
$\tilde{a}$ denotes a sequence of names.
Let $ \fn{P} $ and $ \bn{P} $  denote the sets of free names and bound names occurring in $P$, respectively.
Their definitions are completely standard.
We use $\sigma, \sigma', \sigma_{1}, \ldots$ to range over substitutions.
A substitution is a mapping $[\Subst{x_{1}}{y_{1}}, \ldots ,\Subst{x_{n}}{y_{n}}]$ from names to names.
The application of a substitution on a term $P[\Subst{x_{1}}{y_{1}}, \ldots ,\Subst{x_{n}}{y_{n}}]$ is defined as the result of simultaneously replacing all free occurrences of $y_{i}$ by $x_{i}$ for $i \in \Set{1, \ldots, n}$.
For all names in $\Names \setminus \Set{y_{1}, \ldots, y_{n}}$ the substitution behaves as the identity mapping.
The relation $\step$ as defined in the semantics below defines the reduction steps processes can perform. We write $ P \step P' $ if $ (P, P') \in {\step} $ and call $ P' $ a \emph{derivative} of $ P $.
Let $ \steps $ denote the reflexive and transitive closure of $ \step $.
$ P $ is \emph{divergent} if it has an infinite sequence of steps $ P \step^{\omega} $.
We use \emph{barbs} or \emph{observables} to distinguish between processes with different behaviours. We write $ P\HasBarb{\alpha} $ if $ P $ has a barb $ \alpha $, where the predicate $ \cdot\HasBarb{\cdot} $ can be defined differently for each calculus. Moreover $ P $ has a weak barb $ \alpha $, if $ P $ may reach a process with this barb, \ie $ P\ReachBarb{\alpha} \deff \exists P'\logdot P \steps P' \wedge P'\HasBarb{\alpha} $.

As source calculus we use the following variant of CSP \cite{hoare:78csp}.

\begin{definition}\label{CSPSyntax}
The processes $\ProcCSP$ are given by
$$P \;\mathop{::=}\; \CSPPar{P}{P}{A} \sep \Div \sep \Stop \sep \inchoice{P}{P} \sep P/b \sep f(P) \sep X \sep \mu X \cdot P \sep \textstyle{\ExSum_{i \in \IM}} a_i \rightarrow P$$
	where $X \in \mathcal{X}$ is a process variable, $A \subseteq \Names$, and $ \IM $ is a finite index set.
\end{definition}

$\CSPPar{P}{Q}{A}$ is the parallel composition of $ P $ and $ Q $, where $ P $ and $ Q $ can proceed independently except for actions $ a \in A $, on which they have to synchronise.
$\Div$ describes \emph{divergence}.
$\Stop$ denotes \emph{inaction}.
\emph{Internal choice} $\inchoice{P}{Q}$ reduces to either $ P $ or $ Q $ within a single internal step.
\emph{Concealment} $P/b$ hides an action $ b $ and masks it as $ \tau $.
\emph{Renaming} $f(P)$ for some $f:\Names \to \Names$ extended by $f(\tau)=\tau$ behaves as $ P $, where $ a $ is replaced by $ f(a) $ for all $ a \in \Names $.
\emph{Recursion} $\mu X \cdot P$ describes a process behaving like $P$ with every occurrence of $X$ being replaced by $\mu X \cdot P$.
\emph{External choice} $\ExSum_{i \in \IM} a_i \rightarrow P_i$ offers a selection of one of the \emph{action prefixes} $a_i \rightarrow \cdot $ followed by the corresponding continuation $P_i$, so it may perform any $ a_i $ and then behave like $P_i$. Note that we enforce action prefixes to be syntactically part of an external choice construct.
 As usual, we use $\exchoice{M}{N}$ to denote binary external choice.

The CSP semantics is given by the following rules, using labelled steps $\Trans{\alpha}$ to define $\step$:
\vspace{-0.75em}
\begin{align*}
		\begin{array}{|c|}
			\hline
			\ninfer{\runa{Con$_0$}}{E \Trans{b} E'}{E /b \Trans{\tau} E'/b} \hspace*{2.5em} \ninfer{\runa{Con$_1$}}{E \Trans{\alpha} E' \quad (\alpha \not = b)}{E /b \Trans{\alpha} E'/b}  \hspace*{2.5em} \ninfer{\runa{Ren}}{E \Trans{\alpha} E'}{f(E) \Trans{f(\alpha)} f(E')} \hspace*{2.5em} \ninfer{\runa{Ece}}{M_j \Trans{a} M_j' \quad (j \in \IM)}{\ExSum_{i \in  \IM}M_i \Trans{a} M_j'}\\
			\\
			\ninfer{\runa{Act}}{}{(a \rightarrow E) \Trans{a} E} \hspace*{2.5em} \ninfer{\runa{Rec}}{}{\mu X \cdot E \Trans{\tau} E[\mu X \cdot E/X]}\\
			\\
			\ninfer{\runa{Par$_0$}}{E \Trans{\alpha} E' \quad (\alpha \not \in A)}{\CSPPar{E}{F}{A} \Trans{\alpha} \CSPPar{E'}{F}{A}} \hspace*{1.5em} \ninfer{\runa{Par$_1$}}{F \Trans{\alpha} F' \quad (\alpha \not \in A)}{\CSPPar{E}{F}{A} \Trans{\alpha} \CSPPar{E}{F'}{A}} \hspace*{1.5em} \ninfer{\runa{Par$_2$}}{E \Trans{a} E'\quad F \Trans{a} F' \quad (a \in A)}{\CSPPar{E}{F}{A} \Trans{a} \CSPPar{E'}{F'}{A}} \hspace*{1.5em} \ninfer{\runa{Red}}{P \Trans{\tau} P'}{P \step P'}\\
			\\
			\ninfer{\runa{Div}}{}{\Div \Trans{\tau} \Div} \hspace*{2.5em} \ninfer{\runa{Inc$_0$}}{}{\inchoice{E}{F} \Trans{\tau} E} \hspace*{2.5em} \ninfer{\runa{Inc$_1$}}{}{\inchoice{E}{F} \Trans{\tau} F} 
			\\
			\hline
		\end{array}
\end{align*}
\vspace{-1.0em}\\
A barb of CSP is the possibility of a term, to perform an action, \ie $ P\HasBarb{a} \deff \exists P'\logdot P \Trans{a} P' $.
Following the definition of distributability in \cite{petersNestmannGoltz13} a CSP term $ P $ is distributable into $ P_1, \ldots, P_n $ if $ P_1, \ldots, P_n $ are unguarded subterms of $ P $ such that every action prefix in $ P $ occurs in exactly one of the $ P_1, \ldots, P_n $, where different but equally-named action prefixes are distinguished and unguarded occurrences of $ \mu X \cdot P' $ may result in several copies of $ P' $ within the $ P_1, \ldots, P_n $.

As target calculus we use an asynchronous variant of CCS\cite{CCS} with name-passing and matching.

\begin{definition}\label{def:ccs_syntax}
  The processes $\ProcCCS$ are given by
  \begin{align*}
    P & \;\mathop{::=}\; \Par{P}{P} \sep \Res*{\tilde{c}}{P} \sep \RepInput{\ch}{\tilde{x}}{P} \sep \Input{\ch}{\tilde{x}}{P} \sep \Out{\ch}{\tilde{x}} \sep \Match{\ch}{z}P \sep \Null
  \end{align*}
\end{definition}

$\Par{P}{Q}$ is the parallel composition of $P$ and $Q$, where $P$ and $Q$ can either proceed independently or synchronise on matching channels names.
$\Res*{\tilde{c}}{P}$ restricts the visibility of actions using names in~$\tilde{c}$ to~$P$.
$ \Input{c}{\tilde{x}}{P} $ denotes input on channel $c$.
$ \Out{c}{\tilde{x}} $ is output on channel $c$.
Since there is no continuation, we interpret this calculus as asynchronous.
We use $ \RepInput{\ch}{\tilde{x}}{P} $ to denote \emph{replicated input} on channel $c$ with the continuation $P$.
$ \Match{x}{y}P  $ is the matching operator, if $ x = y $ then $P$ is enabled.
$\Null$ denotes inaction.

The CCS semantics is given by following transition rules:
\vspace{-0.6em}
\begin{align*}
		\begin{array}{|c|}
			\hline
			\ninfer{\runa{Par}}{P \step P'}{\Par{P}{Q} \step \Par{P'}{Q}} \hspace*{2.5em} \ninfer{\runa{Res}}{P \step P'}{\Res*{\tilde{c}}{P} \step \Res*{\tilde{c}}{P'}} \hspace*{2.5em} \ninfer{\runa{Cong}}{P \equiv P' \quad P' \step Q' \quad Q' \equiv Q}{P \step Q}\\
			\\
			\ninfer{\runa{Rep}}{}{\Par{\RepInput{\ch}{\tilde{x}}P}{\Out{\ch}{\tilde{y}}} \step \Par{\RepInput{\ch}{\tilde{x}}P}{P[\tilde{y}/\tilde{x}]}} \hspace*{2.5em} \ninfer{\runa{Com}}{}{\Par{\Out{\ch}{\tilde{y}}}{\Input{\ch}{\tilde{x}}Q} \step \Par{P}{Q[\tilde{y}/\tilde{x}]}}
			\\
			\hline
		\end{array}
\end{align*}
\vspace{-1em}\\
where $ \equiv $ denotes structural congruence given by the rules: $ \Par{P}{0} \equiv P$,
$ \Par{P}{Q} \equiv \Par{Q}{P} $, $ \Par{P}{\left( \Par{Q}{R} \right)} \equiv \Par{\left(\Par{P}{Q} \right)}{R} $, $\Res*{\tilde{a}}{\Null} \equiv \Null $, $\Par{P}{\Res*{\tilde{a}}{Q}} \equiv \Res{\tilde{a}}{\Par{P}{Q}} $ if $\bn{\tilde{a}} \notin \fn{P}$, and $\Match{x}{x}P \equiv P $.
As discussed in \cite{petersNestmannGoltz13}, a CCS term $ P $ is distributable into $ P_1, \ldots, P_n $ if $ P \equiv \Res{\tilde{x}}{P_1 \mid \ldots \mid P_n} $.

\vspace{0.5em}
\noindent
\textbf{Simulation Relations.}\;
The semantics of a process is usually considered modulo some behavioural equivalence.
For many calculi, \emph{the} standard reference equivalence is some form of weak bisimilarity.
In the context of encodings, the source and target language often differ in their relevant obervables, \ie barbs. In this case, it is advantageous to use a variant of reduction bisimilarity.
With Gorla \cite{gorla10}, we add a \emph{success} operator $ \success $ to the syntax of both CSP and CCS. Since $ \success $ cannot be further reduced, the semantics is left unchanged in both cases. The test for the reachability of success is standard in both languages, \ie $ P\hasSuccess \deff \exists P'\logdot P \equiv \success \mid P' $.
To obtain a non-trivial equivalence, we require that the bisimulation respects success and the reachability of barbs.
We use the standard definition of barbs in CSP, \ie action prefixes.
Our encoding function will translate all source terms into closed terms, thus the standard definition of CCS barbs would not provide any information.
Instead we use a notion of translated barb ($ \cdot\ReachBarb{\EncCI{\cdot}} $) that reflects how the encoding function translates source term barbs. Its definition is given in Section~\ref{sec:innerPart}.

\begin{definition}[Bisimulation]
	A relation $ \mathcal{R} \; \subseteq \proc^2 $ is a \emph{(success-sensitive, [translated-]barb-respecting, weak, reduction) bisimulation} if, whenever $ \left( P, Q \right) \in \mathcal{R} $, then:
	\begin{compactitem}
		\item $ P \step P' $ implies $ \exists Q'\logdot Q \steps Q' \wedge \left( P', Q' \right) \in \mathcal{R} $
		\item $ Q \step Q' $ implies $ \exists P'\logdot P \steps P' \wedge \left( P', Q' \right) \in \mathcal{R} $
		\item $ P\reachSuccess $ iff $ Q\reachSuccess $
		\item $ P $ and $ Q $ reach the same (translated) barbs, where we use $ \cdot\ReachBarb{a} $ for CSP and $ \cdot\ReachBarb{\EncCI{a}} $ for CCS
	\end{compactitem}
	Two terms $ P, Q \in \proc $ are \emph{bisimilar}, denoted as $ P \barbBisim Q $, if there exists a bisimulation that relates $ P $ and $ Q $.
\end{definition}

\noindent
We use the symbol $ \barbBisim $ to denote either bisimilarity on our target language CCS or on the disjoint union of CSP and CCS that allows us to describe the relationship between source terms and their translations. In the same way we define a corresponding variant of coupled similarity.

\begin{definition}[Coupled Simulation]
	A relation $ \mathcal{R} \; \subseteq \proc^2 $ is a \emph{(success-sensitive, [translated-]barb-respecting, weak, reduction) coupled simulation} if, whenever $ \left( P, Q \right) \in \mathcal{R} $, then:
	\begin{compactitem}
		\item $ P \step P' $ implies $ \exists Q'\logdot Q \steps Q' \wedge \left( P', Q' \right) \in \mathcal{R} $ and $ \exists Q''\logdot Q \steps Q'' \wedge \left( Q'', P' \right) \in \mathcal{R} $
		\item $ P\reachSuccess $ iff $ Q\reachSuccess $
		\item $ P $ and $ Q $ reach the same (translated) barbs, where we use $ \cdot\ReachBarb{a} $ for CSP and $ \cdot\ReachBarb{\EncCI{a}} $ for CCS
	\end{compactitem}
	Two terms $ P, Q \in \proc $ are \emph{coupled similar}, denoted as $ P \barbCS Q $, if there exists a coupled simulation that relates $ P $ and $ Q $ in both directions.
\end{definition}

\vspace{0.3em}
\noindent
\textbf{Encodings and Quality Criteria.}\;
We consider two different translations from (the above-defined variant of) CSP into (the above-defined variant of) CCS with name passing and matching. 
In this context, we refer to CSP terms as \emph{source terms} $ \procS $ and to CCS terms as \emph{target terms} $ \procT $. 
Encodings often translate single source steps into a sequence or pomset of target steps. We call such a sequence or pomset a \emph{simulation} of the corresponding source term step.
Moreover, we assume for each encoding the existence of a so-called renaming policy $ \varphi $, \ie a mapping of names from the source into vectors of target term names.

To analyse the quality of encodings and to rule out trivial or meaningless encodings, Gorla \cite{gorla10} provide a general framework comprising five quality criteria, which have afterwards been used in many papers.
In addition to our above-mentioned definition of process calculus, whough, Gorla requires the target calculus to be equipped with a notion of behavioural equivalence $ \asymp $ on target terms. 
Its purpose is to describe the `abstract' behaviour of a target process, where `abstract' refers to an observer at the source level. 
In \cite{gorla10}, the equivalence $ \asymp $ is often defined as a barbed equivalence (cf.~\cite{milner.sangiorgi:barbed-bisimulation}) or can be derived directly from the reduction semantics, and it typically is a congruence, at least with respect to parallel composition. 
Bisimilarity and coupled similarity are such relations on CCS terms.
The criteria are:
\begin{compactenum}[(1)]
	\item \emph{Compositionality}: The translation of an operator $ \mathrm{op} $ is the same for all occurrences of that operator in a term, \ie it can be captured by a context $ \context_{\mathrm{op}} $ such that $ \Enc{\mathrm{op}\left( x_1, \ldots, x_n, S_1, \ldots, S_m \right)} = \Context{N}{\mathbf{op}}{x_1, \ldots, x_n, \Enc{S_1}, \ldots, \Enc{S_m}} $ for $ \fn{S_1} \cup \ldots \cup \fn{S_m} = N $.
	\item \emph{Name Invariance}: The encoding does not depend on particular names, \ie for every $ S $ and $ \sigma $, it holds that $ \Enc{\sigma\left( S \right)} \equiv \sigma'\left( \Enc{S} \right) $ if $ \sigma $ is injective and $ \Enc{\sigma\left( S \right)} \asymp \sigma'\left( \Enc{S} \right) $ otherwise, where $ \sigma' $ is such that $ \Renam{\sigma\left( n \right)} = \sigma'\left( \Renam{n} \right) $ for every $ n \in \names $.
	\item \emph{Operational Correspondence}: Every computation of a source term can be simulated by its translation, \ie $ S \stepsS S' $ implies $ \Enc{S} \stepsT \asymp \Enc{S'} $ (completeness), and every computation of a target term corresponds to some computation of the corresponding source term (soundness, compare to Section~\ref{sec:decentral}). 
	\item \emph{Divergence Reflection}: The encoding does not introduce divergence, \ie $ \Enc{S} \stepT^{\omega} $ implies $ S \stepS^{\omega} $.
	\item \emph{Success Sensitiveness}: A source term and its encoding answer the tests for success in exactly the same way, \ie $ S \reachSuccess $ iff $ \Enc{S} \reachSuccess $.
\end{compactenum}

Our encodings will satisfy all of these criteria except for compositionality, because both encodings consists of two layers.
\cite{petersNestmannGoltz13} shows that the above criteria do not ensure that an encoding preserves distribution and proposes an additional criterion for the preservation of distributability.

\begin{definition}[Preservation of Distributability]
	\label{def:distributabilityPreservation}
	
	An encoding $ \enc $ \emph{preserves distributability} if for every $ S $ and for all terms $ S_1, \ldots, S_n $ that are distributable within $ S $ there are some $ T_1, \ldots, T_n $ that are distributable within $ \Enc{S} $ such that $ T_i \asymp \Enc{S_i} $ for all $ 1 \leq i \leq n $.
\end{definition}

\noindent
Here, because of the choice of the source and the target language, an encoding preserves distributability if for each sequence of distributable source term steps their simulations are pairwise distributable. 
In both languages two alternative steps of a term are in \emph{conflict} with each other if---for CSP---they reduce the same action-prefix or---for CCS---they either reduce the same input using two outputs or they reduce the same output using two [replicated] inputs. 
Two alternative steps that are not in conflict are \emph{distributable}.

\section{Translating the CSP Synchronisation Mechanism}
\label{sec:innerPart}

CSP and CCS---or the $ \pi $-calculus---differ fundamentally in their communication and synchronisation mechanisms.
In CSP there is only a single kind of action $ c \rightarrow \cdot $, where $ c $ is a name. 
Synchronisation is implemented by the parallel operator $ \CSPPar{\cdot}{\cdot}{A} $ that in CSP is augmented with a set of names $ A $ containing the names that need to be synchronised at this point. 
By nesting parallel operators arbitrarily many actions on the same name can be synchronised.
In CCS there are two different kinds of actions: inputs $ \In{c}{} $ and outputs $ \Out{c}{} $. Again synchronisation is implemented by the parallel operator, but in CCS only a single input and a single matching output can ever be synchronised within one step.

To encode the CSP communication and synchronisation mechanisms in CCS with name passing we make use of a technique already used in \cite{petersNestmann12, peters12} to translate between different variants of the $ \pi $-calculus. CSP actions are translated into action announcements augmented with a lock indicating, whether the respective action was already used in the simulation of a step. The other operators of CSP are then translated into handlers for these announcements and locks.
The translation of sum combines several actions under the same lock and thus ensures that only one term of the sum can ever be used.
The translation of the parallel operator combines announcements of actions that need to be synchronised into a single announcement under a fresh lock, whose value is determined by the combination of the respective underlying locks at its left and right side. Announcements of actions that do not need to be synchronised are simply forwarded.
A second layer---containing either a centralised or a decentralised coordinator---then triggers and coordinates the simulation of source term steps.

Action announcements are of the form $ \Out{\act}{\ch, \req, \lock, \simu} $: $ \ch $ is the translation of the source term action. $ \req $ is used to trigger the computation of the Boolean value of $ \lock $. The lock $ \lock $ evaluates to $ \top $ as long as the respective translated action was not successfully used in the simulation of a step. $ \simu $ is used to guard the encoded continuation of the respective source term action. In the case of a successful simulation attempt involving this announcement, an output $ \Out{\simu}{\top} $ allows to unguard the encoded source term continuation and ensures that all following evaluations of $ \lock $ return $ \bot $. The message $ \Out{\simu}{\bot} $ indicates an aborted simulation attempt and allows to restore $ \lock $ for later simulation attempts. Once a lock becomes $ \bot $, all request for its computation return $ \bot $.

\vspace{0.3em}
\noindent
\textbf{Abbreviations.}\;
We introduce some abbreviations to simplify the presentation of the encodings. We use
\vspace{-1.9em}
\begin{align*}
	\MatchIn{x}{A}P \deff \textstyle{\prod_{a \in A}} \Match{x}{a}P
\end{align*}
\vspace{-2em}\\
to test, whether an action belongs to the set of synchronised actions in the encoding of the parallel operator.
As already done in \cite{nestmann96, nestmannPierce00} we use Boolean-valued locks to ensure that every translation of an action is only used once to simulate a step.
\emph{Boolean locks} are channels on which only the Boolean values $ \top $ (true) or $ \bot $ (false) are transmitted. 
An unguarded output over a Boolean lock with value $ \top $ represents a positive instantiation of the respective lock, whereas an unguarded output sending $ \bot $ represents a negative instantiation. 
At the receiving end of such a channel, the Boolean value can be used to make a binary decision, which is done here within an \emph{$ \textsc{if} $-construct}.
This construct and according instantiations of locks are implemented as in \cite{nestmann96, nestmannPierce00} using restriction and the order of transmitted values.
\vspace{-0.5em}
\begin{align*}
	\Out{\lock}{\top} \deff \Input{\lock}{t, f}{\Out{t}{}} \quad & \quad\quad \Out{\lock}{\bot} \deff \Input{\lock}{t, f}{\Out{f}{}}\\
	\Input{\lock}{\boolV}{\ITE{\boolV}{P}{Q}} & \deff \Res{t, f}{\Out{\lock}{t, f} \mid \Input{t}{}{P} \mid \Input{f}{}{Q}}
\end{align*}
\vspace{-2em}\\
We observe that the Boolean values $ \top $ and $ \bot $ are realised by a pair of links without parameters. Both cases of the $\textsc{if}$-construct operate as guard for its subterms $ P $ and $ Q $. The renaming policy $ \renam $ reserves the names $ t $ and $ f $ to implement the Boolean values $ \true $ and $ \false $.

\vspace{0.3em}
\noindent
\textbf{The Algorithm.}\;
The encoding functions introduce some fresh names, that are reserved for special purposes. In Table~\ref{tab:resNam} we list the reserved names $ \mathcal{R} $ and provide a hint on their purpose.
\begin{table}[t]
	\begin{tabular}{|c|c|}
		\hline
		reserved names & purpose\\
		\hline
		$ \act $, $ \act' $ & announce the ability to perform an action\\
		$ \ch $, $ \lch $, $ \rch $, $ z $ & (translated) source term channel, channel from the left/right of a parallel operator\\
		$ \lock $, $ \lLock $, $ \rLock $ & lock, lock from the left/right of a parallel operator\\
		$ \lock' $ & re-instantiate a positive sum lock\\
		$ \req $, $ \lreq $, $ \rreq $ & request the computation of the value of a lock\\
		$ \simu $, $ \simu_i $, $ \lsimu $, $ \rsimu $ & simulate a source term step and unguard the corresponding continuations\\
		$ \nextSyn $ & order left announcements for the same channel that need to be synchronised\\
		$ \syn $, $ \syn' $ & distribute right announcements that need to be synchronised\\
		$ \boolV $ & Boolean value ($ \bot $ or $ \top $)\\
		$ \tau $ & fresh name used to announce $ \tau $-steps that result from concealment\\
		$ \once $ & used by the centralised encoding to avoid overlapping simulation attempts\\
		$ \much $ & fresh names used to encode internal choice\\
		$ \rep $ & fresh names used to encode divergence\\
		$ t, f $ & used to encode Boolean values\\
		\hline
	\end{tabular}
	\caption{Reserved Names.}
	\label{tab:resNam}
\end{table}
Moreover we reserve the names $ \Set{ x_i \mid i \in \nat } $ and assume an injective mapping $ \renam': \mathcal{X} \to \Set{ x_i \mid i \in \nat } $ that maps process variables of CSP to distinct names.
The renaming policy $ \renam $ for our encodings is then a function that reserves the names in $ \mathcal{R} \cup \Set{ x_i \mid i \in \nat } $ and translates every source term name into three target term names. More precisely, choose $ \varphi : \names \to \names^3 $ such that:
\begin{compactenum}
	\item No name is mapped onto a reserved name, \ie $ \Renam{n} \cap \left( \mathcal{R} \cup \Set{ x_i \mid i \in \nat } \right) = \emptyset $ for all $ n \in \names $.
	\item No two different names are mapped to overlapping sets of names, \ie $ \Renam{n} \cap \Renam{m} = \emptyset $ for all $ n, m \in \names $ with $ n \neq m $.
\end{compactenum}
We naturally extend the renaming policy to sets of names, \ie $ \Renam{X} \deff \Set{ \Renam{x} \mid x \in X } $ if $ X \subseteq \names $.
Let $ \Proj{\left( x_1, \ldots, x_n \right)}{i} \deff x_i $ denote the projection of a $ n $-tuple to its $ i $th element, if $ 1 \leq i \leq n $. Moreover $ \Proj{X}{i} \deff \Set{ \Proj{x}{i} \mid x \in X } $ for a set $ X $ of $ n $-tuples and $ 1 \leq i \leq n $.

\begin{figure}[htp]
	\begin{align*}
		\EncCI{\CSPPar{P}{Q}{A}} \deff & \res{\act', \Proj{\Renam{A}}{2}, \Proj{\Renam{A}}{3}}\Big(\!\\
			& \quad \Res{\act}{\EncCI{P} \mid \RepInput{\act}{\ch, \tilde{x}}{\left( \MatchIn{\ch}{\Proj{\Renam{A}}{1}} \Out{\Proj{\Renam{\ch}}{2}}{\tilde{x}} \mid \MatchOut{\ch}{\Proj{\Renam{A}}{1}} \Out{\act'}{\ch, \tilde{x}} \right)}}\\
			& \quad \Res{\act}{\EncCI{Q} \mid \RepInput{\act}{\ch, \tilde{x}}{\left( \MatchIn{\ch}{\Proj{\Renam{A}}{1}} \Out{\Proj{\Renam{\ch}}{3}}{\tilde{x}} \mid \MatchOut{\ch}{\Proj{\Renam{A}}{1}} \Out{\act'}{\ch, \tilde{x}} \right)}}\\
			& \quad \mid \textstyle{\prod_{\ch \in A}} \Synch{\ch} \mid \RepInput{\act'}{\tilde{x}}{\Out{\act}{\tilde{x}}}
		\!\Big)
		\\
		\Synch{\ch} \deff & \res{\nextSyn}\Big(\!
			\Out{\nextSyn}{\Proj{\Renam{\ch}}{3}}\\
			& \quad \mid \RepIn{\nextSyn}{\syn}\Big(\In{\Proj{\Renam{\ch}}{2}}{\lreq, \lLock, \lsimu}.\Big( \res{\syn'}\Big(\\
				& \quad\quad \RepInput{\syn}{\rreq, \rLock, \rsimu}{\left( \Res{\req, \lock, \simu}{\Out{\act}{\Proj{\Renam{c}}{1}, \req, \lock, \simu} \mid \Sim} \mid \Out{\syn'}{\rreq, \rLock, \rsimu} \right)}\\
				& \quad\quad \mid \Res{\syn}{\Out{\nextSyn}{\syn} \mid \RepInput{\syn'}{\tilde{x}}{\Out{\syn}{\tilde{x}}}}
			\Big)\!\Big)\!\Big)
		\!\Big)
		\\
		\Sim \deff & \res{\lock'}\Big( \Out{\lock'}{} \mid \RepIn{\lock'}{}.\Big( \In{\req}{}.\Big(\Out{\lreq}{} \mid \In{\lLock}{\boolV}.\Big( \IF{\boolV}\;\THEN{}\Big( \Out{\rreq}{} \mid \In{\rLock}{\boolV}.\Big( \IF{\boolV}\\
				& \quad \quad \THEN{\left( \Out{\lock}{\top} \mid \Input{\simu}{\boolV}{\left( \Out{\lsimu}{\boolV} \mid \Out{\rsimu}{\boolV} \mid \ITE{\boolV}{\RepInput{\req}{}{\Out{\lock}{\bot}}}{\Out{\lock'}{}} \right)} \right)}\\
				& \quad \quad \ELSE{\left( \Out{\lock}{\bot} \mid \Out{\lsimu}{\bot} \right) \mid \RepInput{\req}{}{\Out{\lock}{\bot}}} \Big)\!\Big)\\
			& \quad \ELSE{\left( \Out{\lock}{\bot} \mid \RepInput{\req}{}{\Out{\lock}{\bot}} \right)}\!\Big)\!\Big)\!\Big)\!\Big)
		\\
		\EncCI{\textstyle{\ExSum_{i \in \IM}} \ch_i \to P_i} \deff & \res{\req, \lock, \simu_1, \ldots, \simu_n}\Big( \Input{\req}{}{\Out{\lock}{\top}}\\
			& \mid \textstyle{\prod_{i \in \IM}} \left( \Out{\act}{\Proj{\ch_i}{1}, \req, \lock, \simu_i} \mid \RepIn{\simu_i}{\boolV}.\ITE{\boolV}{\left( \EncCI{P_i} \mid \RepInput{\req}{}{\Out{\lock}{\bot}} \right)}{\Input{\req}{}{\Out{\lock}{\top}}} \right)\!\Big)
		\\
		\EncCI{\left( P \right) / z} \deff & \Res{\act'}{\Res{\act, z}{\EncCI{P} \mid \RepInput{\act}{\ch, \tilde{x}}{\left( \Match{\ch}{z}\Out{\act'}{\tau, \tilde{x}} \mid \MissMatch{\ch}{z}\Out{\act'}{\ch, \tilde{x}} \right)}} \mid \RepInput{\act'}{\tilde{x}}{\Out{\act}{\tilde{x}}}}
		\\
		\EncCI{f(P)} \deff & \res{\act'}\big( \res{\act, z}\big( \EncCI{P} \mid \RepIn{\act}{\ch, \tilde{x}}\big( \textstyle{\prod_{z/x \in f}} \Match{\ch}{\Proj{\Renam{x}}{1}}\Out{\act'}{\Proj{\Renam{z}}{1}, \tilde{x}}\\
		& \mid \MatchOut{\ch}{\Dom{f}}\Out{\act'}{\ch, \tilde{x}} \big)\!\big) \mid \RepInput{\act'}{\tilde{x}}{\Out{\act}{\tilde{x}}}\big)
		\\
		\EncCI{\Div} \deff & \Res{\rep}{\Out{\rep}{} \mid \RepInput{\rep}{}{\Out{\rep}{}}}
		\\ 
		\EncCI{\mu X \cdot P} \deff & \Res{\renam'(X)}{\Out{\renam'(X)}{} \mid \RepInput{\renam'(X)}{}{\EncCI{P}}}
		\\
		\EncCI{X} \deff & \Out{\renam'(X)}{}
		\\
		\EncCI{\inchoice{P}{Q}} \deff & \Res{\much}{\Input{\much}{}{\EncCI{P}} \mid \Input{\much}{}{\EncCI{Q}} \mid \Out{\much}{}}
		\\ 
		\EncCI{\Stop} \deff & \Null
		\\ 
		\EncCI{\success} \deff & \success
	\end{align*}
	where $ \notin \Proj{\Renam{A}}{1} $ is short for $ \in \left( \fn{P} \cup \fn{Q} \right) \setminus \Proj{\Renam{A}}{1} $, $ \notin \Dom{f} $ is short for $ \in \fn{P} \setminus \Dom{f} $, and $ \neq z $ is short for $ \in \fn{P} \setminus \Set{ z } $.
	\caption{An encoding from CSP into CCS with value passing (inner part).}
	\label{fig:innerEncoding}
\end{figure}
The inner part of our two encodings is presented in Figure~\ref{fig:innerEncoding}. The most complex case is the translation of the parallel operator $ \EncCI{\CSPPar{P}{Q}{A}} $ that is based on the following four steps:
\begin{description}
	\item[Step 1:] Action announcements for channels $ \ch \notin A $\\
		In the case of actions on channels $ \ch \notin A $---that do not need to be synchronised here---the encoding of the parallel operator acts like a forwarder and transfers action announcements of both its subtrees further up in the parallel tree.
		Two different restrictions of the channel for action announcements $ \act $ from the left side $ \EncCI{P} $ and the right side $ \EncCI{Q} $, allow to trace action announcements back to their origin as it is necessary in the following case.
		In the present case we use $ \act' $ to bridge the action announcement over the restrictions on $ \act $.
	\item[Step 2:] Action announcements for channels $ \ch \in A $\\
		Actions $ \ch \in A $ need to be synchronised, \ie can be performed only if both sides of the parallel operator cooperate on this action. Simulating this kind of synchronisation is the main purpose of the encoding of the parallel operator.
		The renaming policy $ \renam $ translates each source term name into three target term names. The first target term name is used as reference to the original source term name and transferred in announcements. The other two names are used to simulate the synchronisation of the parallel operator in CSP. Announcements from the left are translated to outputs on the respective second name and announcements from the right to the respective third name. Restriction ensures that these outputs can only be computed by the current parallel operator encoding. The translations of the announcements into different outputs for different source term names allows us to treat announcements of different names concurrently using the term $ \Synch{\ch} $, where $ \ch $ is a source term name.
	\item[Step 3:] The term $ \Synch{\ch} $\\
		In $ \Synch{\ch} $ all announcements for the same source term name $ \ch $ from the left are ordered in order to combine each left and each right announcement on the same name. Several such announcements may result from underlying parallel operators, sums with similar summands, and junk left over from already simulated source term steps. For each left announcement a fresh instance of $ \syn $ is generated and restricted. The names $ \syn $ and $ \syn' $ are used to transfer right announcements to the respective next left announcement, where $ \syn' $ is used to bridge over the restriction on $ \syn $. This way each right announcement will eventually be transferred to each left announcement on the same name. Note that this kind of forwarding is not done concurrently but in the source language a term $ \CSPPar{P}{Q}{A} $ also cannot perform two steps on the same name $ \ch \in A $ concurrently. After combining a left and a right announcement on the same source term name a fresh set of auxiliary variables $ \req, \lock, \simu $ is generated and a corresponding announcement is transmitted. The term $ \Sim $ reacts to requests regarding this announcement and is used to simulate a step on the synchronised action.
	\item[Step 4:] The term $ \Sim $\\
		If a request reaches $ \Sim $ it starts questioning the left and the right side. First the left side is requested to compute the current value of the lock of the action. Only if $ \top $ is returned, the right side is requested to compute its lock as well. This avoids deadlocks that would result from blindly requesting the computation of locks in the decentralised encoding. If the locks of both sides are still valid the fresh lock $ \lock $ returns $ \top $ else $ \bot $ is returned. For each case $ \Sim $ ensures that subsequently requests will obtain an answer by looping with $ \Out{\lock'}{} $ or returning $ \bot $ to all requests, respectively. The messages $ \Out{\lsimu}{\bot} $ and $ \Out{\rsimu}{\bot} $ cause the respective underlying subterms on the left and the right side to do the same, whereas $ \Out{\lsimu}{\top} $ and $ \Out{\rsimu}{\top} $ cause the unguarding of encoded continuations as result of a successful simulation of a source term synchronisation step.
\end{description}

\vspace{0.3em}
\noindent
\textbf{Basic Properties and Translated Observables.}\;
The protocol introduced by the encoding function in Figure~\ref{fig:innerEncoding} (and its outer parts introduced later) simulates a single source term step by a sequence of target term steps. Most of these steps are merely pre- and post-processing steps, \ie they do not participate in decisions regarding the simulation of conflicting source term steps but only prepare and complete simulations. Accordingly we distinguish between \emph{auxiliary steps}---that are pre- and post-processing steps---and \emph{simulation steps}---that mark a point of no return by deciding which source term step is simulated. Note that the points of no return and thus the definition of auxiliary and simulation steps is different in the two variants of our encoding.

Auxiliary steps do not influence the choice of source terms steps that are simulated. Moreover they operate on restricted channels, \ie are unobservable. Accordingly they do not change the state of the target term modulo the considered reference relations $ \barbBisim $ and $ \barbCS $. We introduce some auxiliary lemmata to support this claim.

The encoding $ \EncCI{\cdot} $ translates source term barbs $ \ch $ into free announcements with $ \Proj{\Renam{\ch}}{1} $ as first value and a lock $ \lock $ as third value that computes to $ \top $. The two coordinators, \ie outer encodings, we introduce later, restrict the free $ \act $-channel of $ \EncCI{\cdot} $.

\begin{definition}[Translated Barbs]
	Let $ T \in \procT $ such that $ \exists S\logdot \EncCI{S} \stepsT T $, $ \exists S\logdot \EncCO{S} \stepsT T $, or $ \exists S\logdot \EncDO{S} \stepsT T $.
	$ T $ has a translated barb $ \ch $, denoted by $ T\HasBarb{\EncCI{\ch}} $, if
	\begin{compactitem}
		\item there is an unguarded output $ \Out{\act}{\Proj{\Renam{\ch}}{1}, \req, \lock, \simu} $---on a free channel $ \act $ in the case of $ \EncCI{\cdot} $ or the outermost variant of $ \act $ in the case of the later introduced encodings $ \EncCO{\cdot} $ and $ \EncDO{\cdot} $---in $ T $ or
		\item such an announcement was consumed to unguard an $ \textsc{if} $-construct testing $ \lock $ and this construct is still not resolved in $ T $
	\end{compactitem}
	such that all locks that are necessary to instantiate $ \lock $ are positively instantiated.
\end{definition}

Analysing the encoding function in Figure~\ref{fig:innerEncoding} we observe that an encoded source term has a translated barb iff the corresponding source term has the corresponding source term barb.

\begin{obs}
	For all $ S \in \procS $, it holds $ S\HasBarb{\ch} $ iff $ \EncCI{S}\HasBarb{\EncCI{\ch}} $.
	\label{obs:transBarbs}
\end{obs}

All instances of success in the translation result from success in the source. More precisely the only way to obtain $ \success $ in the translation is by $ \EncCI{\success} \deff \success $.

\begin{obs}
	For all $ S \in \procS $, it holds $ S\hasSuccess $ iff $ \EncCI{S}\hasSuccess $.
	\label{obs:success}
\end{obs}

The encoding propagates announcements through the translated parallel structure. In the translation of parallel operators it combines all left and right announcements \wrt to the same channel name, if this channel needs to be synchronised. Therefore we copy announcements.
We use locks carrying a Boolean value to indicate whether an announcement was already used to simulate a source term step. These locks carry $ \top $ in the beginning and are swapped to $ \bot $ as soon as the announcement was used. In each state there is at most one positive instantiation of each lock and as soon as a lock is instantiated negatively it never becomes positive again.

\begin{lemma}
	Let $ T \in \procT $ such that $ \exists S\logdot \EncCI{S} \stepsT T $. Then for each variant $ l $ of the names $ \lock, \lLock, \rLock $
	\begin{compactenum}
		\item there is at most one positive instantiation of $ l $ in $ T $,
		\item if there is a positive instantiation of $ l $ in $ T $ then there is no other instantiation of $ l $ in $ T $,
		\item if there is a negative instantiation of $ l $ in $ T $ then no derivative of $ T $ contains a positive instantiation of $ l $.
	\end{compactenum}
	\label{lem:sumLocks}
\end{lemma}

\section{The Centralised Encoding}
\label{sec:central}

Figure~\ref{fig:innerEncoding} describes how to translate CSP actions into announcements augmented with locks and how the other operators are translated to either forward or combine these announcements and locks. With that $ \EncCI{\cdot} $ provides the basic machinery of our encoding from CSP into CCS with name passing and matching. However it does not allow to simulate any source term step. Therefore we need a second (outer) layer that triggers and coordinates the simulation of source term steps. We consider two ways to implement this coordinator: a centralised and a decentralised coordinator. The centralised coordinator is depicted in Figure~\ref{fig:centralised}.

\begin{figure}
	\begin{align*}
		\EncCO{P} \deff & \Res{\act, \once}{\EncCI{P} \mid \Out{\once}{} \mid\RepInput{\once}{}{\Input{\act}{\ch, \req, \lock, \simu}{\left( \Out{\req}{} \mid \Input{\lock}{\boolV}{\left( \Out{\once}{} \mid \IF{\boolV}\;\THEN{\Out{\simu}{\top}} \right)} \right)}}}
	\end{align*}
	\caption{A \textbf{centralised} encoding from CSP into CCS with value passing.}
	\label{fig:centralised}
\end{figure}

The channel $ \once $ is used to ensure that simulation attempts of different source term steps cannot overlap each other. For each simulation attempt exactly one announcement is consumed. The coordinator then triggers the computation of the respective lock that was transmitted in the announcement. This request for the computation of the lock is propagated along the parallel structure induced by the translations of parallel operators until---in the leafs---encodings of sums are reached. There the request for the computation yields the transmission of the current value of the respective lock. While being transmitted back to the top of the tree, different locks that refer to synchronisation in the source terms are combined. If the computation of the lock results with $ \top $ at the top of the tree, the respective source term step is simulated. Else the encoding aborts the simulation attempt and restores the consumed informations about the values of the respective locks. In both cases a new instance of $ \Out{\once}{} $ allows to start the next simulation attempt. Accordingly only some post-processing steps can overlap with a new simulation attempt.

As we prove below, the points of no return in the centralised encoding can result from the consumption of action announcements by the outer encoding in Figure~\ref{fig:centralised} if the corresponding lock computes to $ \top $. Moreover the encoding of internal choice and divergence introduces simulation steps, namely all steps on variants of the channels $ \much $, $ \rep $, and $ \renam'(X) $. All remaining steps of the centralised encoding are auxiliary.

\begin{definition}[Auxiliary and Simulation Steps]
	A step $ T \stepT T' $ such that $ \exists S \in \procS\logdot \EncCO{S} \stepsT T $ is called a \emph{simulation step}, denoted by $ T \sStepT T' $, if $ T \step T' $ is a step on the outermost channel $ \act $ and the computation of the value of the received lock $ \lock $ will return $ \top $ or it is a step on a variant of $ \much $, $ \rep $, or $ \renam'(X) $.
	
	Else the step $ T \stepT T' $ is called an \emph{auxiliary step}, denoted by $ T \aStepT T' $.
	\label{def:auxStepsCentral}
\end{definition}

\noindent
Let $ \aStepsT $ denote the reflexive and transitive closure of $ \aStepT $ and let $ \sStepsT \deff \aStepsT \sStepT \aStepsT $.
Auxiliary steps do not change the state modulo $ \barbBisim $.

\begin{lemma}
	$ T \aStepT T' $ implies $ T \barbBisim T' $ for all target terms $ T, T' $.
	\label{lem:auxStepsCentral}
\end{lemma}

By distinguishing auxiliary and simulation steps, we can prove a condition stronger than operational correspondence, namely that each source term step is simulated by exactly one simulation step.

\begin{lemma}
	For all $ S, S' $, it holds $ S \stepS S' $ iff $ \exists T\logdot \EncCO{S} \sStepsT T \wedge \EncCO{S'} \barbBisim T $.
	\label{lem:sourceVsSimStep}
\end{lemma}

\noindent
This direct correspondence between source term steps and the points of no return of their translation allows us to prove a variant of operational correspondence that is significantly stricter than the variant proposed in \cite{gorla10}.

\begin{definition}[Operational Correspondence]
	$ $\\
	An encoding $ \enc: \procS \to \procT $ is \emph{operationally corresponding} \wrt $ \barbBisim \; \subseteq \procT^2 $ if it is:
	\begin{compactitem}
		\item[\; Complete:] $ \forall S, S' \logdot S \stepsS S' $ implies $ \exists T \logdot \EncCO{S} \stepsT T \wedge \EncCO{S'} \barbBisim T $
		\item[\; Sound:] $ \forall S, T \logdot \EncCO{S} \stepsT T $ implies $ \exists S' \logdot S \stepsS S' \wedge \EncCO{S'} \barbBisim T $
	\end{compactitem}
\end{definition}

\noindent
The `if'-part of Lemma~\ref{lem:sourceVsSimStep} implies operational completeness \wrt $ \barbBisim $ and the `only-if'-part contains the main argument for operational soundness \wrt $ \barbBisim $. Hence $ \encCO $ is operationally corresponding \wrt to $ \barbBisim $.

\begin{theorem}
	The encoding $ \encCO $ is operationally corresponding \wrt to $ \barbBisim $.
	\label{thm:operationalCorrespondenceCentral}
\end{theorem}

To obtain divergence reflection we show that there is no infinite sequence of only auxiliary steps.
Then divergence reflection follows from the combination of this fact and Lemma~\ref{lem:sourceVsSimStep}.

\begin{theorem}
	The encoding $ \encCO $ reflects divergence.
	\label{thm:divergenceReflectionCentral}
\end{theorem}

The encoding function ensures that $ \EncCO{S} $ has an unguarded occurrence of $ \success $ iff $ S $ has such an unguarded occurrence. Operational correspondence ensures that $ S $ and $ \EncCO{S} $ also answer the question for the reachability of $ \success $ in the same way.

\begin{theorem}
	The encoding $ \encCO $ is success sensitive.
	\label{thm:successSensitivenessCentral}
\end{theorem}

In a similar way we can prove that a source term reaches a barb iff its translation reaches the respective translated barb.

\begin{theorem}
	For all $ S, \ch $, it holds $ S\ReachBarb{\ch} $ iff $ \EncCO{S}\ReachBarb{\EncCI{\ch}} $.
	\label{thm:respectsBarbsCentral}
\end{theorem}

As proved in \cite{petersGlabbeek15}, Theorem~\ref{thm:operationalCorrespondenceCentral}, the fact that $ \barbBisim $ is success sensitive and respects (translated) barbs, Theorem~\ref{thm:successSensitivenessCentral}, and Theorem~\ref{thm:respectsBarbsCentral} imply that for all $ S $ it holds $ S $ and $ \EncCO{S} $ are (success sensitive, (translated) barb respecting, weak, reduction) bisimilar, \ie $ S \barbBisim \EncCO{S} $.
Bisimilarity is a strong relation between source terms and their translation. On the other hand, because of efficiency, distributability preserving encodings are more interesting.
Because of $ \once $ the encoding $ \encCO $ obviously does not preserve distributability. As discussed in \cite{parrowCoupled92} bisimulation often forbids distributed encodings. Instead they propose coupled simulation as a relation that still provides a strong connection between source terms and their translations but is more flexible. Following the approach in \cite{parrowCoupled92} we consider a decentralised coordinator next.

\section{The Decentralised Encoding}
\label{sec:decentral}

\begin{figure}
	\begin{align*}
		\EncDO{P} \deff & \Res{\act}{\EncCI{P} \mid \RepInput{\act}{\ch, \req, \lock, \simu}{\left( \Out{\req}{} \mid \Input{\lock}{\boolV}{\IF{\boolV}\;\THEN{\Out{\simu}{\top}}} \right)}}
	\end{align*}
	\caption{A \textbf{decentralised} encoding from CSP into CCS with value passing.}
	\label{fig:decentralised}
\end{figure}

Figure~\ref{fig:decentralised} presents a decentralised variant of the coordinator in Figure~\ref{fig:centralised}.
The only difference between the centralised and the decentralised version of the coordinator is that the latter can request to check different locks concurrently. Technically $ \encCO $ and $ \encDO $ differ only by the use of $ \once $. As a consequence the steps of different simulation attempts can overlap and even (pre-processing) steps of simulations of conflicting source term steps can interleave to a certain degree. Because of this effect, $ \encDO $ does not satisfy the version of operational correspondence used above for $ \encCO $, but $ \encDO $ satisfies weak operational correspondence that was proposed in \cite{gorla10} as part of a set of quality criteria.

Since several announcements can be processed concurrently by the decentralised coordinator, here all consumptions of announcements are auxiliary steps. Instead the consumption of positive instantiations of locks can mark a point of no return. In contrast to $ \encCO $ not every point of no return in $ \encDO $ unambiguously marks a simulation of a single source term step, because in contrast to $ \encCO $ the encoding $ \encDO $ introduces \emph{partial commitments} \cite{peters12,petersNestmann12}.

Consider the example $ E = \CSPPar{\left( \exchoice{o \to P_1}{p \to P_2} \right)}{\left( \exchoice{o \to P_3}{\exchoice{p \to P_4}{q \to P_5}} \right)}{\Set{ o, p }} $.

\noindent
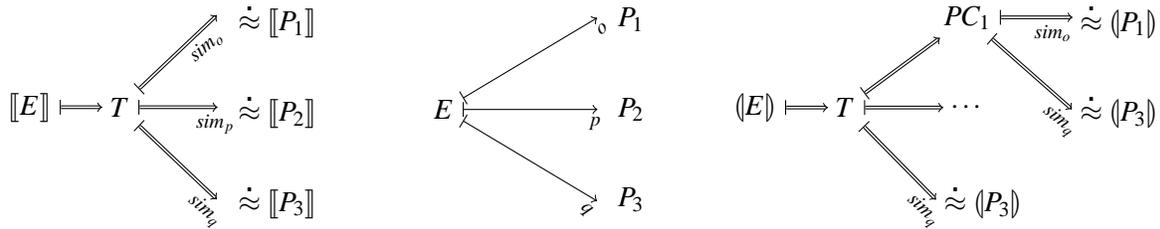
  \begin{minipage}[c]{0.3\textwidth-2pt}
      \begin{tikzpicture}[auto,node distance=1.2cm]
        \node (E)                        {$ \EncCO{E} $};
        \node (T)    [right of=E]        {$ T $};
        
        \node (T2)    [right=1cm of T]        {$\hspace{1ex} \barbBisim \EncCO{P_2} $};
        \node (T1)    [above of=T2]        {$\hspace{1ex} \barbBisim \EncCO{P_1} $};
        \node (T3)    [below of=T2]        {$\hspace{1ex} \barbBisim \EncCO{P_3} $};

        \draw[|->,shorten >=-1pt,shorten <=-0.5pt] (E) -- (T);
        \draw[double] (E) -- (T);
        \draw[|->,shorten >=-1pt,shorten <=-0.5pt] (T) -- (T1.west) node [near end, below, rotate=40, scale = 0.7] {$\emph{sim}_o$};
        \draw[double] (T) -- (T1.west);
        \draw[|->,shorten >=-1pt,shorten <=-0.5pt] (T) -- (T2.west) node [at end, below, scale = 0.7] {$\emph{sim}_p$};
        \draw[double] (T) -- (T2.west);
        \draw[|->,shorten >=-1pt,shorten <=-0.5pt] (T) -- (T3.west) node [at end, below, rotate=-40, scale = 0.7] {$\emph{sim}_q$};
        \draw[double] (T) -- (T3.west);
      \end{tikzpicture}
  \end{minipage}
  \begin{minipage}[c]{0.3\textwidth-2pt}
    \begin{center}
      \begin{tikzpicture}[auto,node distance=1.2cm]
        \node (T)                        {$ E $};
        \node (d1)    [right of=T]        {};
        \node (T2)    [right of=d1]        {$\hspace{1ex} P_2$};
        \node (T1)    [above of=T2]        {$\hspace{1ex} P_1$};
        \node (T3)    [below of=T2]        {$\hspace{1ex} P_3$};

        \draw[|->,shorten >=-1pt,shorten <=-0.5pt] (T) -- (T1.west) node [at end, below, rotate=40, scale = 0.7] {$o$};
        \draw[|->,shorten >=-1pt,shorten <=-0.5pt] (T) -- (T2.west) node [at end, below, scale = 0.7] {$p$};
        \draw[|->,shorten >=-1pt,shorten <=-0.5pt] (T) -- (T3.west) node [at end, below,rotate=-40, scale = 0.7] {$q$};
      \end{tikzpicture}
      \end{center}
  \end{minipage}
  \begin{minipage}[c]{0.4\textwidth-2pt}
      \begin{tikzpicture}[auto,node distance=1.2cm]
        \node (E)                        {$ \EncDO{E} $};
        \node (T)    [right of=E]        {$ T $};
        \node (P2)    [right=1cm of T]        {$\cdots$};
        \node (P1)    [above of=P2]        {$PC_1$};
        \node (T12)    [right=1cm of P2]        {$\barbBisim \EncDO{P_{3}}$};
        \node (T11)    [above of=T12]        {$\barbBisim \EncDO{P_{1}}$};
        \node (T3)    [below of=P2]        {$\makebox[\widthof{$PC_{1}$}][l]{$\barbBisim \EncDO{P_{3}}$}$};

        \draw[|->,shorten >=-1pt,shorten <=-0.5pt] (E) -- (T);
        \draw[double] (E) -- (T);
        \draw[|->,shorten >=-1pt,shorten <=-0.5pt] (T) -- (P1);
        \draw[double] (T) -- (P1);
        \draw[|->,shorten >=-1pt,shorten <=-0.5pt] (T) -- (P2);
        \draw[double] (T) -- (P2);
        \draw[|->,shorten >=-1pt,shorten <=-0.5pt] (P1) -- (T11.west) node [near end, below, scale = 0.7] {$\emph{sim}_o$};
        \draw[double] (P1) -- (T11.west);
        \draw[|->,shorten >=-1pt,shorten <=-0.5pt] (P1) -- (T12.west) node [at end, below, rotate=-40, scale = 0.7] {$\emph{sim}_q$};
        \draw[double] (P1) -- (T12.west);
        \draw[|->,shorten >=-1pt,shorten <=-0.5pt] (T) -- (T3.west) node [at end, below, rotate=-43, scale = 0.7] {$\emph{sim}_q$};
        \draw[double] (T) -- (T3.west);
      \end{tikzpicture}
  \end{minipage}

  In the example, two sides of a parallel operator have to synchronise on either action $p$, or action $o$, or action $q$ happens without synchronisation.
  In the centralised encoding $\EncCO{E}$ the use of $ \once $ ensures that different simulation attempts cannot overlap. Thus, only after finishing the simulation of a source term step, the simulation of another source term step can be invoked. As a consequence each state reachable from encoded source terms can unambiguously be mapped to a single state of the source term. This allows us to use a stronger version of operational correspondence and, thus, to prove that source terms and their translations are bisimilar. The corresponding 1-to-1 correspondence between source terms and their translations is visualised by the first two graphs above, where $ T \barbBisim \EncCO{E} $.

  The decentralised encoding $\EncDO{E}$ introduces partial commitments.
  Assume the translation of a source term that offers several alternative ways to be reduced. Then some encodings---as our decentralised one---do not always decide on which of the source term steps should be simulated next. More precisely a partial commitment refers to a state reachable from the translation of a source term in that already some possible simulations of source term steps are ruled out, but there is still more than a single possibility left.
  
  In the decentralised encoding announcements can be processed concurrently and parts of different simulation attempts can interleave. The only blocking part of the decentralised encoding are conflicting attempts to consume the same positive instantiation of a lock.
  In the presented example above there are two locks; one for each side of the parallel operator. The simulations of the step on $ o $ and $ p $ need both of these locks, whereas to simulate the step on $ q $ only a positive instantiation of the right lock needs to be consumed.
  By consuming the positive instantiation of the left lock in an attempt to simulate the step on $ o $, the simulation of the step on $ p $ is ruled out, but the simulation of the step on $ q $ is still possible. Since either the simulation of the step on $ o $ or the simulation of the step on $ q $ succeeds, the simulation of the step on $ p $ is not only blocked but ruled out. But the consumption of the instantiation of the left lock does not unambiguously decide between the remaining two simulations. The intermediate state that results from consuming the instantiation of the left lock and represents a partial commitment is visualised in the right graph above by the state $ PC_1 $.
  
  Partial commitments forbid a 1-to-1 mapping between the states of  a source term and its translations by a bisimulation. But, as shown in \cite{parrowCoupled92}, partial commitments do not forbid to relate source terms and their translations by coupled similarity.

Whether the consumption of a positive instantiation of a lock is an auxiliary step---does not change the state of the term modulo $ \barbBisim $---, is a partial commitment, or unambiguously marks a simulation of a single source term step depends on the surrounding term, \ie cannot be determined without the context. For simplicity we consider all steps that reduce a positive instantiation of a lock as simulation steps.
Also steps on variants of the channels $ \much $, $ \rep $, and $ \renam'(X) $ are simulation steps, because they unambiguously mark a simulation of a single source term step. All remaining steps of the decentralised encoding are auxiliary.

\begin{definition}[Auxiliary and Simulation Steps]
	A step $ T \stepT T' $ such that $ \exists S \in \procS\logdot \EncDO{S} \stepsT T $ is called a \emph{simulation step}, denoted by $ T \sStepT T' $, if $ T \step T' $ reduces a positive instantiation of a lock or is a step on a variant of $ \much $, $ \rep $, or $ \renam'(X) $.
	
	Else the step $ T \stepT T' $ is called an \emph{auxiliary step}, denoted by $ T \aStepT T' $.
	\label{def:auxStepsDecentral}
\end{definition}

\noindent
Again let $ \aStepsT $ denote the reflexive and transitive closure of $ \aStepT $ and let $ \sStepsT \deff \aStepsT \sStepT \aStepsT $.
Since auxiliary steps do not introduce partial commitments, they do not change the state modulo $ \barbBisim $. The proof of this lemma is very similar to the centralised case.

\begin{lemma}
	$ T \aStepT T' $ implies $ T \barbBisim T' $ for all target terms $ T, T' $.
	\label{lem:auxStepsDecentral}
\end{lemma}

In contrast to the centralised encoding, the simulation of a source term step in the decentralised encoding can require more than a single simulation step and a single simulation step not unambiguously refers to the simulation of a particular source term step. The partial commitments described above forbid operational correspondence, but the weaker variant proposed in \cite{gorla10} is satisfied. We call this variant weak operational correspondence.

\begin{definition}[Weak Operational Correspondence]
	$ $\\
	An encoding $ \enc: \procS \to \procT $ is \emph{weakly operationally corresponding} \wrt $ \barbCS \; \subseteq \procT^2 $ if it is:
	\begin{compactitem}
		\item[\; Complete:] $ \forall S, S' \logdot S \stepsS S' $ implies $ \exists T \logdot \EncDO{S} \stepsT T \wedge \EncDO{S'} \barbCS T $
		\item[\; Weakly Sound:] $ \forall S, T \logdot \EncDO{S} \stepsT T $ implies $ \exists S', T' \logdot S \stepsS S' \wedge T \stepsT T' \wedge \EncDO{S'} \barbCS T' $
	\end{compactitem}
\end{definition}

The only difference to operational correspondence is the weaker variant of soundness that allows for $ T $ to be an intermediate state that does not need to be related to a source term directly. Instead there has to be a way from $ T $ to some $ T' $ such that $ T' $ is related to a source term.

\begin{theorem}
	The encoding $ \encDO $ is weakly operational corresponding \wrt to $ \barbBisim $.
	\label{thm:operationalCorrespondenceDecentral}
\end{theorem}

As in the encoding $ \encCO $, there is no infinite sequence of only auxiliary steps in $ \EncDO{S} $.
Moreover each simulation of a source term requires only finitely many simulation steps (to consume the respective positive instantiations of locks). Thus $ \encDO $ reflects divergence.

\begin{theorem}
	The encoding $ \encDO $ reflects divergence.
	\label{thm:divergenceReflectionDecentral}
\end{theorem}

The encoding function ensures that $ \EncDO{S} $ has an unguarded occurrence of $ \success $ iff $ S $ has such an unguarded occurrence. Operational correspondence again ensures that $ S $ and $ \EncDO{S} $ also answer the question for the reachability of $ \success $ in the same way.

\begin{theorem}
	The encoding $ \encDO $ is success sensitive.
	\label{thm:successSensitivenessDecentral}
\end{theorem}

Similarly, a source term reaches a barb iff its translation reaches the respective translated barb.

\begin{theorem}
	For all $ S, \ch $, it holds $ S\ReachBarb{\ch} $ iff $ \EncDO{S}\ReachBarb{\EncCI{\ch}} $.
	\label{thm:respectsBarbsDecentral}
\end{theorem}

Weak operational correspondence does not suffice to establish a bisimulation between source terms and their translations.
But, as proved in \cite{petersGlabbeek15}, Theorem~\ref{thm:operationalCorrespondenceDecentral}, the fact that $ \barbBisim $ is success sensitive and respects (translated) observables, Theorem~\ref{thm:successSensitivenessDecentral}, and Theorem~\ref{thm:respectsBarbsDecentral} imply that $ \forall S\logdot S $ and $ \EncCO{S} $ are (success sensitive, (translated) barbs respecting, weak, reduction) coupled similar, \ie $ S \barbCS \EncDO{S} $.

It remains to show, that $ \encDO $ indeed preserves distributability. Therefore we prove that all blocking parts of the encoding $ \encDO $ refer to simulations of conflicting source term steps.

\begin{theorem}
	The encoding $ \encDO $ preserves distributability.
	\label{thm:distributability}
\end{theorem}

\section{Conclusions}
\label{sec:conclusion}
 
We introduced two encodings from CSP into asynchronous CCS with name passing and matching.
As in \cite{parrowCoupled92} we had to encode the multiway synchronisation mechanism of CSP into binary communications and, similarly to \cite{parrowCoupled92}, we did so first using a centralised controller that was then modified into a decentralised controller.
By doing so we were able to transfer the observations of \cite{parrowCoupled92} to the present case:
\begin{compactenum}
	\item The centralised solution allows to prove a stronger connection between source terms and their translations, namely by bisimilarity. Our decentralised solution does not relate source terms and their translations that strongly and we doubt that any decentralised solution can do so.
	\item Nonetheless, decentralised solutions are possible as presented by the second encoding and they still relate source terms and their translations in an interesting way, namely by coupled similarity.
\end{compactenum}
Thus as in \cite{parrowCoupled92} we observed a trade-off between \emph{centralised} but \emph{bisimilar} solutions on the one-hand side and \emph{decentralised} but only \emph{coupled similar} solutions on the other side.

More technically we showed here instead a trade-off between centralised but \emph{operationally corresponding} solutions on the one-hand side and \emph{weakly operationally corresponding} but decentralised solutions on the other side.
The mutual connection between operational correspondence and bisimilarity as well as between weak operational correspondence and coupled similarity is proved in \cite{petersGlabbeek15}.

Both encodings make strict use of the renaming policy and translate into closed terms.
Hence the criterion \emph{name invariance} is trivially satisfied in both cases.
Moreover we showed that both encodings are \emph{success-sensitive}, \emph{reflect divergence}, and even \emph{respect barbs} \wrt to the standard source term (CSP) barbs and a notion of translated barbs on the target.
The centralised encoding $ \encCO $ additionally satisfies a variant of \emph{operational correspondence} that is stricter than the variant proposed in \cite{gorla10}.
The decentralised encoding $ \encDO $ satisfies \emph{weak operational correspondence} as proposed in \cite{gorla10} and \emph{distributability preservation} as proposed in \cite{petersNestmannGoltz13}.
Thus both encodings satisfy all of the criteria proposed in \cite{gorla10} except for compositionality.
However in both cases the inner part is obviously compositional and the outer part only adds a fixed context.

\providecommand{\thisvolume}[2][]{this volume of EPTCS}
\def\opa{}
\bibliography{cspToCcs}

\begin{thebibliography}{10}
\providecommand{\bibitemdeclare}[2]{}
\providecommand{\surnamestart}{}
\providecommand{\surnameend}{}
\providecommand{\urlprefix}{Available at }
\providecommand{\url}[1]{\texttt{#1}}
\providecommand{\href}[2]{\texttt{#2}}
\providecommand{\urlalt}[2]{\href{#1}{#2}}
\providecommand{\doi}[1]{doi:\urlalt{http://dx.doi.org/#1}{#1}}
\providecommand{\bibinfo}[2]{#2}

\bibitemdeclare{article}{Baeten:2005:BHP:1085667.1085669}
\bibitem{Baeten:2005:BHP:1085667.1085669}
\bibinfo{author}{J.~C.~M. \surnamestart Baeten\surnameend}
  (\bibinfo{year}{2005}): \emph{\bibinfo{title}{{A Brief History of Process
  Algebra}}}.
\newblock {\sl \bibinfo{journal}{Theor. Comput. Sci.}}
  \bibinfo{volume}{335}(\bibinfo{number}{2-3}), pp. \bibinfo{pages}{131--146},
  \doi{10.1016/j.tcs.2004.07.036}.

\bibitemdeclare{inproceedings}{DBLP:conf/pstv/Brinksma85}
\bibitem{DBLP:conf/pstv/Brinksma85}
\bibinfo{author}{E.~\surnamestart Brinksma\surnameend} (\bibinfo{year}{1985}):
  \emph{\bibinfo{title}{A tutorial on {LOTOS}}}.
\newblock In: {\sl \bibinfo{booktitle}{Proc. of PSTV}}, pp.
  \bibinfo{pages}{171--194}.

\bibitemdeclare{inproceedings}{DBLP:conf/icalp/Brookes83}
\bibitem{DBLP:conf/icalp/Brookes83}
\bibinfo{author}{S.~D. \surnamestart Brookes\surnameend}
  (\bibinfo{year}{1983}): \emph{\bibinfo{title}{On the Relationship of {CCS}
  and {CSP}}}.
\newblock In: {\sl \bibinfo{booktitle}{Proc. of ICALP}}, {\sl
  \bibinfo{series}{LNCS}} \bibinfo{volume}{154}, pp. \bibinfo{pages}{83--96},
  \doi{10.1007/BFb0036899}.

\bibitemdeclare{inproceedings}{7092761}
\bibitem{7092761}
\bibinfo{author}{H.~\surnamestart Evrard\surnameend} \&
  \bibinfo{author}{F.~\surnamestart Lang\surnameend} (\bibinfo{year}{2015}):
  \emph{\bibinfo{title}{Automatic Distributed Code Generation from Formal
  Models of Asynchronous Concurrent Processes}}.
\newblock In: {\sl \bibinfo{booktitle}{Proc. of PDP}},
  \bibinfo{publisher}{IEEE}, pp. \bibinfo{pages}{459--466},
  \doi{10.1109/PDP.2015.96}.

\bibitemdeclare{inproceedings}{DBLP:journals/corr/abs-1208-2750}
\bibitem{DBLP:journals/corr/abs-1208-2750}
\bibinfo{author}{R.~\surnamestart van Glabbeek\surnameend}
  (\bibinfo{year}{2012}): \emph{\bibinfo{title}{{Musings on Encodings and
  Expressiveness}}}.
\newblock In: {\sl \bibinfo{booktitle}{Proc. of EXPRESS/SOS}}, {\sl
  \bibinfo{series}{EPTCS}}~\bibinfo{volume}{89}, pp. \bibinfo{pages}{81--98},
  \doi{10.4204/EPTCS.89.7}.

\bibitemdeclare{article}{gorla10}
\bibitem{gorla10}
\bibinfo{author}{D.~\surnamestart Gorla\surnameend} (\bibinfo{year}{2010}):
  \emph{\bibinfo{title}{{Towards a Unified Approach to Encodability and
  Separation Results for Process Calculi}}}.
\newblock {\sl \bibinfo{journal}{Information and Computation}}
  \bibinfo{volume}{208}(\bibinfo{number}{9}), pp. \bibinfo{pages}{1031--1053},
  \doi{10.1016/j.ic.2010.05.002}.

\bibitemdeclare{techreport}{hatzelTechRep15}
\bibitem{hatzelTechRep15}
\bibinfo{author}{M.~\surnamestart Hatzel\surnameend},
  \bibinfo{author}{C.~\surnamestart Wagner\surnameend},
  \bibinfo{author}{K.~\surnamestart Peters\surnameend} \&
  \bibinfo{author}{U.~\surnamestart Nestmann\surnameend}
  (\bibinfo{year}{2015}): \emph{\bibinfo{title}{{Encoding CSP into CCS
  (Extended Version)}}}.
\newblock \bibinfo{type}{Technical Report}.
\newblock \urlprefix\url{http://arxiv.org/abs/1508.01127}.

\bibitemdeclare{article}{hoare:78csp}
\bibitem{hoare:78csp}
\bibinfo{author}{C.~A.~R. \surnamestart Hoare\surnameend}
  (\bibinfo{year}{1978}): \emph{\bibinfo{title}{{Communicating Sequential
  Processes}}}.
\newblock {\sl \bibinfo{journal}{Communications of the ACM}}
  \bibinfo{volume}{21}(\bibinfo{number}{8}), pp. \bibinfo{pages}{666--677},
  \doi{10.1145/359576.359585}.

\bibitemdeclare{article}{Hoare2006209}
\bibitem{Hoare2006209}
\bibinfo{author}{C.~A.~R. \surnamestart Hoare\surnameend}
  (\bibinfo{year}{2006}): \emph{\bibinfo{title}{Why ever CSP?}}
\newblock {\sl \bibinfo{journal}{Electronic Notes in Theoretical Computer
  Science}} \bibinfo{volume}{162}(\bibinfo{number}{0}), pp.
  \bibinfo{pages}{209--215}, \doi{10.1016/j.entcs.2006.01.031}.

\bibitemdeclare{article}{Lanese200655}
\bibitem{Lanese200655}
\bibinfo{author}{I.~\surnamestart Lanese\surnameend} \&
  \bibinfo{author}{U.~\surnamestart Montanari\surnameend}
  (\bibinfo{year}{2006}): \emph{\bibinfo{title}{Hoare vs Milner: Comparing
  Synchronizations in a Graphical Framework With Mobility}}.
\newblock {\sl \bibinfo{journal}{Electronic Notes in Theoretical Computer
  Science}} \bibinfo{volume}{154}(\bibinfo{number}{2}), pp. \bibinfo{pages}{55
  -- 72}, \doi{10.1016/j.entcs.2005.03.032}.

\bibitemdeclare{book}{CCS}
\bibitem{CCS}
\bibinfo{author}{R.~\surnamestart Milner\surnameend} (\bibinfo{year}{1980}):
  \emph{\bibinfo{title}{A calculus of communicating systems}}.
\newblock \bibinfo{publisher}{Springer}, \doi{10.1007/3-540-10235-3}.

\bibitemdeclare{inproceedings}{DBLP:conf/ifip/Milner86}
\bibitem{DBLP:conf/ifip/Milner86}
\bibinfo{author}{R.~\surnamestart Milner\surnameend} (\bibinfo{year}{1986}):
  \emph{\bibinfo{title}{Process Constructors and Interpretations (Invited
  Paper)}}.
\newblock In: {\sl \bibinfo{booktitle}{{IFIP} Congress}}, pp.
  \bibinfo{pages}{507--514}.

\bibitemdeclare{inproceedings}{milner.sangiorgi:barbed-bisimulation}
\bibitem{milner.sangiorgi:barbed-bisimulation}
\bibinfo{author}{R.~\surnamestart Milner\surnameend} \&
  \bibinfo{author}{D.~\surnamestart Sangiorgi\surnameend}
  (\bibinfo{year}{1992}): \emph{\bibinfo{title}{Barbed Bisimulation}}.
\newblock In: {\sl \bibinfo{booktitle}{Proc. of ICALP}}, {\sl
  \bibinfo{series}{LNCS}} \bibinfo{volume}{623}, pp. \bibinfo{pages}{685--695},
  \doi{10.1007/3-540-55719-9\_114}.

\bibitemdeclare{phdthesis}{nestmann96}
\bibitem{nestmann96}
\bibinfo{author}{U.~\surnamestart Nestmann\surnameend} (\bibinfo{year}{1996}):
  \emph{\bibinfo{title}{{On Determinacy and Nondeterminacy in Concurrent
  Programming}}}.
\newblock Ph.D. thesis, \bibinfo{school}{Universit{\"a}t
  Erlangen-N{\"u}rnberg}.

\bibitemdeclare{article}{nestmannPierce00}
\bibitem{nestmannPierce00}
\bibinfo{author}{U.~\surnamestart Nestmann\surnameend} \&
  \bibinfo{author}{B.~C. \surnamestart Pierce\surnameend}
  (\bibinfo{year}{2000}): \emph{\bibinfo{title}{{Decoding Choice Encodings}}}.
\newblock {\sl \bibinfo{journal}{Information and Computation}}
  \bibinfo{volume}{163}(\bibinfo{number}{1}), pp. \bibinfo{pages}{1--59},
  \doi{10.1006/inco.2000.2868}.

\bibitemdeclare{inproceedings}{parrowCoupled92}
\bibitem{parrowCoupled92}
\bibinfo{author}{J.~\surnamestart Parrow\surnameend} \&
  \bibinfo{author}{P.~\surnamestart Sj\"odin\surnameend}
  (\bibinfo{year}{1992}): \emph{\bibinfo{title}{Multiway synchronization
  verified with coupled simulation}}.
\newblock In: {\sl \bibinfo{booktitle}{Proc. of CONCUR}}, {\sl
  \bibinfo{series}{LNCS}} \bibinfo{volume}{630},
  \bibinfo{organization}{Springer}, pp. \bibinfo{pages}{518--533},
  \doi{10.1007/bfb0084813}.

\bibitemdeclare{phdthesis}{peters12}
\bibitem{peters12}
\bibinfo{author}{K.~\surnamestart Peters\surnameend} (\bibinfo{year}{2012}):
  \emph{\bibinfo{title}{{Translational Expressiveness}}}.
\newblock Ph.D. thesis, \bibinfo{school}{TU Berlin}.

\bibitemdeclare{inproceedings}{petersGlabbeek15}
\bibitem{petersGlabbeek15}
\bibinfo{author}{K.~\surnamestart Peters\surnameend} \&
  \bibinfo{author}{R.~\surnamestart van Glabbeek\surnameend}
  (\bibinfo{year}{2015}): \emph{\bibinfo{title}{{Analysing and Comparing
  Encodability Criteria}}}.
\newblock In: {\sl \bibinfo{booktitle}{Proc. of EXPRESS/SOS}},
  \bibinfo{series}{\thisvolume[EPTCS]{6}}.

\bibitemdeclare{conference}{petersNestmann12}
\bibitem{petersNestmann12}
\bibinfo{author}{K.~\surnamestart Peters\surnameend} \&
  \bibinfo{author}{U.~\surnamestart Nestmann\surnameend}
  (\bibinfo{year}{2012}): \emph{\bibinfo{title}{{Is it a ``Good'' Encoding of
  Mixed Choice?}}}
\newblock In: {\sl \bibinfo{booktitle}{Proc. of FoSSaCS}}, {\sl
  \bibinfo{series}{LNCS}} \bibinfo{volume}{7213}, pp.
  \bibinfo{pages}{210--224}, \doi{10.1007/978-3-642-28729-9\_14}.

\bibitemdeclare{incollection}{petersNestmannGoltz13}
\bibitem{petersNestmannGoltz13}
\bibinfo{author}{K.~\surnamestart Peters\surnameend},
  \bibinfo{author}{U.~\surnamestart Nestmann\surnameend} \&
  \bibinfo{author}{U.~\surnamestart Goltz\surnameend} (\bibinfo{year}{2013}):
  \emph{\bibinfo{title}{{On Distributability in Process Calculi}}}.
\newblock In: {\sl \bibinfo{booktitle}{Proc. of ESOP}}, {\sl
  \bibinfo{series}{LNCS}} \bibinfo{volume}{7792},
  \bibinfo{publisher}{Springer}, pp. \bibinfo{pages}{310--329},
  \doi{10.1007/978-3-642-37036-6\_18}.

\bibitemdeclare{phdthesis}{sjodin:phd}
\bibitem{sjodin:phd}
\bibinfo{author}{P.~\surnamestart Sj{\"o}din\surnameend}
  (\bibinfo{year}{1991}): \emph{\bibinfo{title}{{From LOTOS Specifications to
  Distributed Implementations}}}.
\newblock Ph.D. thesis, \bibinfo{school}{Department of Computer Science,
  Uppsala University}.
\newblock \bibinfo{note}{Available as Report DoCS 91/31}.

\end{thebibliography}

\end{document}